\documentclass[a4paper,11pt]{article}
\pdfoutput=1 

\usepackage{jinstpub} 

\usepackage{circuitikz}
\usepackage{siunitx}
\usepackage{lineno}
\usepackage{xcolor}

\graphicspath{{img/}}

\newcommand{\Vth}{\ensuremath{V_{\tiny\mbox{th}}}}
\newcommand{\sigmaTot}{\ensuremath{\sigma_{\mbox{\tiny tot}}}}
\newcommand{\sigmaDrift}{\ensuremath{\sigma_{\mbox{\tiny T}}}}
\newcommand{\sigmaOther}{\ensuremath{\sigma_{\mbox{\tiny other}}}}
\newcommand{\sigmaInitial}{\ensuremath{\sigma_{\mbox{\tiny o}}}}

\newcommand{\Iavg}{\ensuremath{I_{\mbox{\tiny avg}}}}
\newcommand{\QofR}{\ensuremath{\tilde{Q}(r)}}
\newcommand{\Iconst}{\ensuremath{I_{\mbox{o}}}}
\newcommand{\Mohm}{\ensuremath{\mbox{M}\Omega}}

\title{\boldmath First operation of a multi-channel Q-Pix prototype: measuring transverse electron diffusion in a gas time projection chamber}

\author[a,1]{N.~Hoch,\note{Co-corresponding authors.}}
\author[b,1]{O.~Seidel,}
\author[b]{V.A.~Chirayath,}
\author[b]{A.B.~Enriquez,}
\author[c]{E.~Gramellini,}
\author[c]{R.~Guenette,}
\author[a]{I.W.~Jaidee,}
\author[d,e]{K.~Keefe,}
\author[d]{S.~Kohani,}
\author[c,f]{S.~Kubota,}
\author[b]{H.~Mahdy,}
\author[b,e]{A.D.~McDonald,}
\author[g]{Y.~Mei,}
\author[g]{P.~Miao,}
\author[h]{F.M.~Newcomer,}
\author[b]{D.~Nygren,}
\author[b]{I.~Parmaksiz,}
\author[b]{M.~Rooks,}
\author[b]{I.~Tzoka,}
\author[a]{W.-Z.~Wei,}
\author[b]{J.~Asaadi,}
\author[a]{J.B.R.~Battat}

\affiliation[a]{Department of Physics and Astronomy, Wellesley College, \\ Wellesley, MA 02481, USA}
\affiliation[b]{Department of Physics, University of Texas at Arlington, \\ Arlington, TX 76019, USA}
\affiliation[c]{Department of Physics and Astronomy, University of Manchester,\\Manchester, UK}
\affiliation[d]{Department of Physics and Astronomy, University of Hawaii, \\ Honolulu, HI 96822, USA}
\affiliation[e]{IF Scientific, \\ Arlington, TX 76018, USA}
\affiliation[f]{Department of Physics, Harvard University,\\Cambridge, MA 02138, USA}
\affiliation[g]{Lawrence Berkeley National Laboratory,\\Berkeley, CA 94720, USA}
\affiliation[h]{Department of Physics and Astronomy, University of Pennsylvania\\Philadelphia, PA 19104, USA}

\emailAdd{ehoch@wellesley.edu}
\emailAdd{olivia.seidel@uta.edu}

\abstract{We report measurements of the transverse diffusion of electrons in P-10 gas (90\% Ar, 10\% CH$_4$) in a laboratory-scale time projection chamber (TPC) utilizing a novel pixelated signal capture and digitization technique known as Q-Pix. The Q-Pix method incorporates a precision switched integrating transimpedance amplifier whose output is compared to a threshold voltage. Upon reaching the threshold, a comparator sends a ‘reset’ signal, initiating a discharge of the integrating capacitor. The time difference between successive resets is inversely proportional to the average current at the pixel in that time interval, and the number of resets is directly proportional to the total collected charge. We developed a 16-channel Q-Pix prototype fabricated from commercial off-the-shelf components and coupled them to 16 concentric annular anode electrodes to measure the spatial extent of the electron swarm that reaches the anode after drifting through the uniform field of the TPC. The swarm is produced at a gold photocathode using pulsed UV light. The measured transverse diffusion agrees with simulations in PyBoltz across a range of operating pressures (200--1500
\,Torr). These results demonstrate that a Q-Pix readout can successfully reconstruct the ionization topology in a TPC.}

\keywords{Charge transport and multiplication in gas, Time projection Chambers (TPC), Data acquisition circuits, Electronic detector readout concepts (gas, liquid)}

\begin{document}
\maketitle
\flushbottom

\section{Introduction}\label{sec:intro}
A 3D pixelated readout for a kiloton-scale liquid-noble time-projection chamber (TPC) requires low-power, low-threshold microelectronics that can operate in a cryogenic environment, and that are scalable to large channel count for high spatial resolution.  To address these challenges, a readout concept called Q-Pix has been proposed \cite{Nygren:2018rbl}. Q-Pix measures the time it takes to accumulate a fixed amount of ionization charge. The detector signal current can be reconstructed from these measured time intervals. Here, we report results from the first prototype of a multi-channel Q-Pix readout that was used to measure the transverse diffusion of electrons in a gas TPC. Our result gives confidence in the ability of Q-Pix to reconstruct ionization events while satisfying the stringent conditions for a fine-grain pixelated read out.

\subsection{Pixel readouts for Time Projection Chambers} 
Large-scale, fine granularity ionization charge readout architectures are a continuing area of research and development for multi-ton-scale noble element detectors. The application and ubiquity of noble element detectors in the fields of high energy physics \cite{ARNEODO199795,Acciarri_2017,Acciarri_2020,Anderson:2012vc,Zhang_2011,WILLIS1974221,Abi_2020,DUNE:2020cqd}, medical imaging 
\cite{Hernandez:2020fpm,GRIGNON2007142,CHEPEL1997427}, and rare event searches \cite{Acciarri_2010,AMAUDRUZ20191,Collaboration_2009,Aprile_2012,Akerib_2014,Anton_2019,_lvarez_2013} stems from their compelling attributes in detecting particle interactions. Charged particles traversing noble element detectors deposit energy in the form of scintillation light and ionization charge. The choice of whether to apply an external electric field to collect ionization electrons depends on the specific application. Doing so creates the familiar TPC, which offers the advantage of reconstructing the three-dimensional trajectory of charged particles~\cite{Nygren_1974}. TPCs provide fully active and uniform tracking detectors with calorimetric reconstruction capabilities.

A fruitful charge readout strategy for liquid argon TPCs (LArTPCs) employs multiple consecutive planes of wires to measure two of three spatial coordinates and deposited energy. 
The third dimension of the event can be obtained by correlating signals across wire planes. This method was used for ICARUS~\cite{antonello2015operation} and MicroBooNE~\cite{acciarri2017design}, as well as many other recent experiments~\cite{Anderson:2012vc,LArIAT:2019kzd,Bian:2015qka}. It was also adopted as a baseline configuration for the Deep Underground Neutrino Experiment (DUNE) far detector~\cite{acciarri2016long}. Wire readouts, however, have an intrinsic limitation in resolving ambiguities in dense, complex topologies, which poses challenges for event reconstruction. Novel event reconstruction techniques help mitigate these difficulties~\cite{Qian:2018qbv, MicroBooNE:2020vry, MicroBooNE:2020jgj, MicroBooNE:2021zul, MicroBooNE:2021ojx}, 
but using pixel-based readouts instead would eliminate this problem and provide an intrinsic 3D readout with improved signal detection efficiency and background rejection~\cite{Adams:2019uqx}. Additionally, pixel detectors can achieve lower energy thresholds than wire readouts due to the smaller capacitance of a pixel compared with that of a long wire~\cite[section 3.10]{Spieler2005-pv}.

Pixelating a large-scale LArTPC requires a readout with very low power dissipation for operation in a cryogenic environment (on the order of 100\,$\mu$W/channel for a pixel granularity of a few millimeters), and the ability to scale the readout to large channel counts (two to three orders of magnitude more channels than a wire readout with comparable spatial resolution). Several approaches are under exploration to achieve this goal. The LArPix group, notably, has demonstrated the feasibility of a scalable, low-power pixelated readout by operating a custom application-specific integrated circuit (ASIC) in LArTPCs~\cite{Dwyer:2018phu}. The LArPix technology has been adopted for use in the DUNE near detector~\cite{DUNE:2021tad}. While the LArPix and Q-Pix concept share similar charge integration circuits that trigger upon a threshold being met, they differ fundamentally in their data readout and clocking schemes. LArPix digitizes the signal current vs. time, while Q-Pix registers the time required to accumulate a fixed amount of signal charge. The timestamp is then transmitted off chip, and the time between resets is used to reconstruct the charge seen. LArPix, designed for high data throughput, excels in environments with high event rates, like those anticipated for the DUNE near detector. QPix, prioritizing low data rate readout, is well suited for scenarios with lower event rates, similar to what is expected in the DUNE far detector. A different pixel readout method, Q-Pix, has also been proposed~\cite{Nygren:2018rbl}. This work presents results from a prototype Q-Pix implementation.

\subsection{Q-Pix}
\begin{figure}[tbp]
 \centering  
 \includegraphics[width=0.5\textwidth]{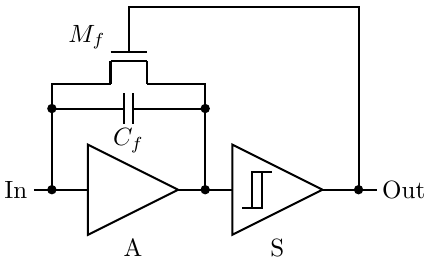}
 \caption{\label{fig:qpix} Schematic of the Q-Pix front-end, which consists of a charge-sensitive amplifier (A) that integrates signal current on a feedback capacitor $C_f$ until a threshold on a Schmitt trigger (S) is met, at which point a reset pulse drains charge from the feedback capacitor and the cycle repeats. Figure from Ref.~\cite{Nygren:2018rbl}.}
 \end{figure}

At its core, Q-Pix is a Charge-Integrate/Reset (CIR) circuit composed of a charge-sensitive amplifier that integrates signal current on a feedback capacitor until a threshold on a Schmitt trigger (regenerative comparator) is met (see figure~\ref{fig:qpix}). Upon reaching this threshold, the Schmitt trigger initiates a rapid ``reset'', draining the feedback capacitor and restoring the circuit to a stable baseline, leaving it ready to begin a new cycle. The time of the ``reset'' transition is captured by reading a local clock. This mode of operation transforms the basic quantum of information for each pixel from the traditional ``charge per unit time'' to ``time per unit charge,'' where $\Delta Q$ is the fixed amount of charge required to initiate a reset, and the time between resets is referred to as the Reset Time Difference (RTD). 
Signal waveforms can be reconstructed from RTDs through the inverse correlation between average input current and  RTD ($\Iavg \propto 1/\mathrm{RTD}$), where \Iavg{} represents the average current over a time interval $\Delta T=\mbox{RTD}$.  
In other words $\Delta Q =  \int I(t) \,\mathrm{d}t = \Iavg \cdot \Delta T$.

A Q-Pix readout of a large-scale LArTPC would bring all of the benefits of pixels relative to wire readouts described previously. In addition, the low noise of Q-Pix will provide enhanced sensitivity to low energy events. For example, a recent study showed a 25\% improvement in the reconstruction efficiency of 5\,MeV supernova neutrinos, and an increase in the supernova burst triggering efficiency by a factor of 2 for events with only five neutrino interactions~\cite{Q-Pix:2022zjm}. Furthermore, the data burden of Q-Pix is vastly reduced relative to a wire readout since each pixel is self-triggered and only times of resets, and not waveforms, are saved. As an example, the data rate from radiogenic backgrounds in a 10\,kton DUNE-like far detector instrumented with Q-Pix would be $10^6$ times smaller than for the same detector with a DUNE-like wire readout~\cite{Q-Pix:2022zjm}.

A full Q-Pix system will eventually be implemented with a dedicated ASIC. To demonstrate the Q-Pix functionality in hardware, a series of Q-Pix prototypes have been developed using commercially available off-the-shelf (COTS) components. One of these prototypes has successfully demonstrated the ability to reconstruct a time-varying current waveform using a reset threshold of $<3000$ electrons, as desired for the eventual ASIC system \cite{Miao:2023ivo}. Here, we present results from a different COTS prototype called the Simplified Analog Q-Pix (SAQ). The SAQ is a 16-channel COTS front-end coupled to an FPGA-based back-end that records the reset timestamps for offline event reconstruction. To demonstrate the functionality of the SAQ, we couple it to a gas-based TPC and measure the transverse diffusion of electrons generated by a pulsed UV light source incident on a photocathode.

Section~\ref{sec:diff} provides an overview of the theory of electron diffusion in gases and details the expected transverse diffusion of electrons in P-10 gas under our operating conditions. Section~\ref{sec:saq} describes our Q-Pix COTS implementation and the associated gas-based TPC. Our measurements of transverse diffusion are presented in section~\ref{sec:measurements}. We conclude by discussing the implications of the successful operation of the SAQ detector in section~\ref{sec:conclusion}.

\section{Electron diffusion in gas}\label{sec:diff}
In an electron-drift gas TPC, a uniform drift field of magnitude $E$ drives ionization electrons toward the readout plane. Along the way, the drifting electrons diffuse -- that is, they scatter elastically or inelastically with gas molecules in the TPC, causing the spatial distribution of the electron swarm to grow. For a point-like electron source at the origin, and for diffusion that is symmetric in all three dimensions, the probability of finding an electron at a perpendicular distance $x$ away from the drift axis after a time $t$ can be expressed as \cite{Sauli2022,Rolandi2008}:
\begin{equation}\label{eq:diffgaus}
    p(x) = \frac{1}{\sqrt{4\pi Dt}} e^{-x^2/4Dt},
\end{equation}
where $D$ is the diffusion constant. The diffusion parallel to the drift field (longitudinal direction) can differ from that perpendicular to the drift direction (transverse direction), especially at high fields, so gases may have different longitudinal and transverse diffusion constants. Here, we focus on transverse diffusion, and write the standard deviation of the Gaussian distribution in the transverse direction as
\begin{equation}\label{eq:sigmaTFirst}
    \sigmaDrift^2 = 2 D_T t,
\end{equation}
where the subscript $T$ indicates the transverse direction.

The transverse diffusion constant $D_T$ depends on the characteristic energy of the ionization electrons, which, 
itself results from competing effects: the kinetic energy lost through collisions with the gas, and the energy gained from the drift field. In an elastic collision, the electron loses a very small fraction of its kinetic energy (${\sim}10^{-4}$), as determined kinematically by the ratio of the electron and gas molecule masses. If, however, the gas molecule has a low threshold for internal excitation, then the electron can excite a transition in the molecule at the expense of a sizable fraction (${\sim} 10^{-1}$) of its kinetic energy. Adding a gas with accessible internal states to a TPC can therefore help reduce diffusion and preserve track geometry. Take, for example, P-10 gas, which was used in this study. P-10 is a 90:10 mixture of argon and methane (CH$_4$). The energy required to excite argon is 11.5\,eV, while for methane it is more than an order of magnitude lower: 0.03\,eV. Therefore, the characteristic electron energy (and resulting diffusion) in CH$_4$ (a ``cold electron gas") is much lower than in pure argon (a ``hot electron gas"). The P-10 mixture preserves the favorable aspects of an argon target while suppressing diffusion thanks to the methane additive. The effect of accessible internal energy states on the characteristic  energy of ionization electrons also explains why trace impurities like water vapor in a TPC may have an outsized impact on electron transport properties. 

\begin{figure}
    \centering
    \includegraphics[width=0.95\textwidth]{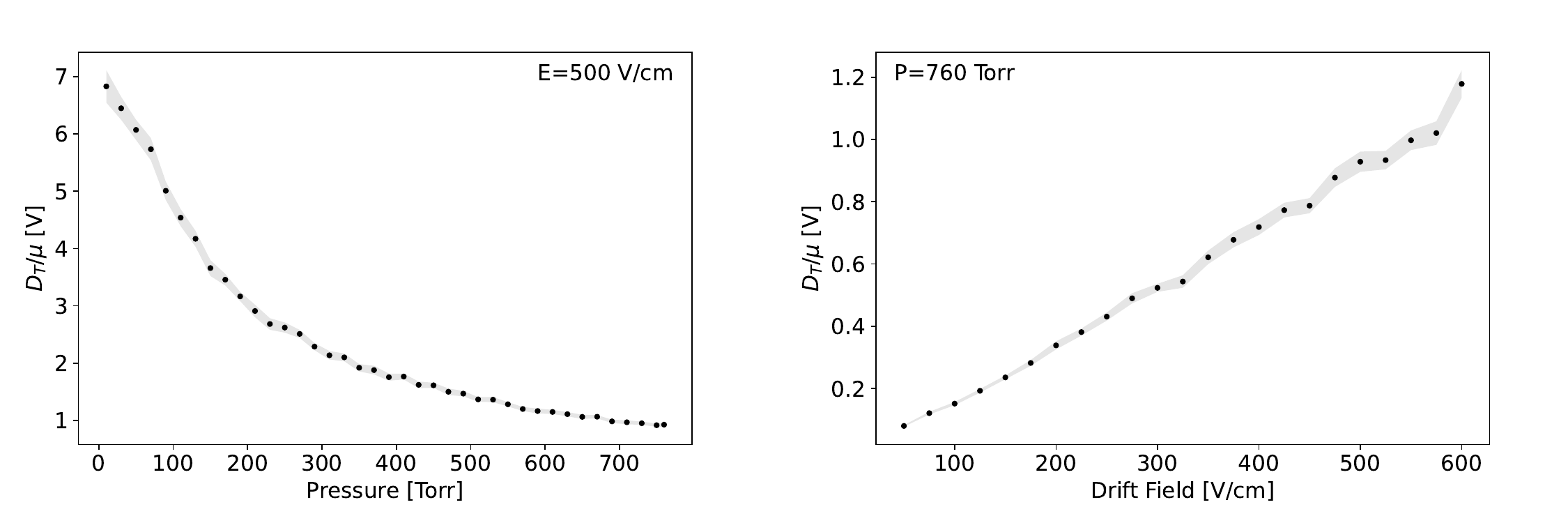}
    \caption{Dependence of $D_T/\mu$, the ratio of the transverse diffusion constant to mobility, for electrons in pure P-10 gas as a function of pressure (left) and drift field (right), computed by PyBoltz~\cite{AlAtoum2020}. Gray bands indicate the 1-$\sigma$ uncertainty intervals.}
    \label{fig:dbymu}
\end{figure}

The electron drift velocity $v_d$ is typically specified as the product of the electric field and the electron mobility $\mu$ via
\begin{equation}\label{eq:vd}
    v_d = \mu E,
\end{equation}
where $\mu$ itself depends on $E$. The diffusion constant and electron mobility are related through
\begin{equation}\label{eq:dbymu}
    \frac{D}{\mu} = \frac{\varepsilon_k}{e},
\end{equation}
where $\varepsilon_k$ is the characteristic electron energy, and $e$ is the electric charge. Thermal electrons have $\varepsilon_k = kT$, where $T$ is the temperature and $k$ is Boltzmann's constant. In that case, Eq.~\ref{eq:dbymu} becomes the familiar Nernst-Townsend formula $D/\mu = kT/e$.
The diffusion of electrons in gas typically exceeds the thermal limit, because $\varepsilon_k \gg kT$. Figure~\ref{fig:dbymu} shows $D_T/\mu$ for electrons in pure P-10 gas as a function of gas pressure and drift field computed using PyBoltz, a Monte Carlo code for simulating electron transport through different gas mixtures relevant to particle detectors \cite{AlAtoum2020}. In the parameter range shown, $D_T/\mu$ is nearly linear with drift field, meaning that the diffusion \sigmaDrift{} is independent of field, since $\sigmaDrift^2\propto (D_T/\mu)/E$. In contrast \sigmaDrift{}  decreases as gas pressure increases (for a fixed drift field).

Using Eqns.~\ref{eq:vd} and \ref{eq:dbymu}, the transverse diffusion (Eq.~\ref{eq:sigmaTFirst}) can be expressed in terms of more convenient experimental parameters such as the drift distance $z=v_d t$ and the drift electric field $E$
\begin{equation}\label{eq:sigmaT}
    \sigmaDrift = \sqrt{\frac{2 D_T z}{\mu E}} = \sqrt{\frac{2\varepsilon_k z}{e E}}.
\end{equation}
This gives the familiar result that diffusion grows with the square root of drift distance 
for a constant electric field (as is the case for a drift field in a TPC).

In practice, the measured transverse dimension of the electron cloud at the anode plane \sigmaTot{} will have contributions from other factors in addition to \sigmaDrift{}. For example, the initial charge cloud at the photocathode may not be point-like, but instead have a width \sigmaInitial{}. The measured spatial spread of the charge in the transverse direction can also be influenced by the instrument response function. For example, in the present case, the TPC elements, such as a gas amplification device or the pixelization scale of the anode plane can contribute an amount \sigmaOther{} to the measured transverse dimension of the charge cloud. If these contributions are independent then they add in quadrature and the measured transverse width \sigmaTot{} is given by:
\begin{equation}\label{eq:sigmaTot}
    \sigmaTot^2 = \sigmaDrift^2 + \sigmaInitial^2 +  \sigmaOther^2.
\end{equation}

Measurements of diffusion using different drift lengths within the same apparatus allow for the determination of \sigmaDrift{}, as it is the only term that varies with drift length. In our apparatus, on the other hand, the drift length was fixed, but independent constraints on \sigmaInitial{} enable us to report $\sqrt{
\sigmaDrift^2+\sigmaOther^2}$.

\section{Experimental design and the Simplified Analog Q-Pix readout}\label{sec:saq}
We demonstrate the operation of our multi-channel Q-Pix prototype by measuring the transverse diffusion of electrons in a gas TPC. A cloud of photoelectrons is generated at the cathode of the TPC using pulsed UV light.
As the electrons drift through the TPC, they diffuse, as described in section~\ref{sec:diff}. The spatial extent of this cloud in the transverse direction is subsequently measured at the anode readout plane.  

\begin{figure}[t]
 \centering 
 \includegraphics[width=0.95\textwidth]{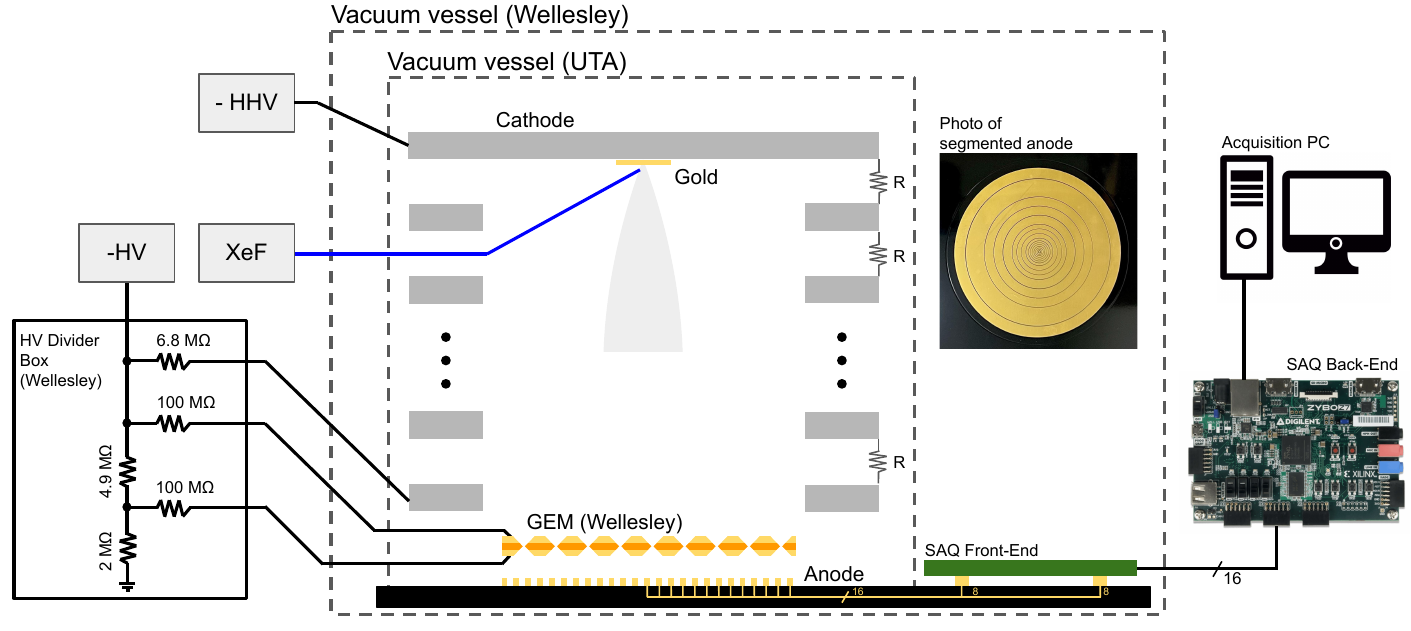}
 \caption{\label{fig:bothSystems} System diagram for the TPC and SAQ electronics. A xenon flashlamp (XeF) delivers UV light via optical fiber to a gold foil attached to the cathode to produce a cloud of electrons that drift toward the anode (indicated by the light gray cone emerging from the gold). A uniform drift field is established by a set of field shaping electrodes connected by resistors $R$. The anode is segmented into 16 concentric annuli (see inset photo), each coupled to a separate readout channel on the SAQ front end. The front-end PCB (green) mates to the anode board (black) via two rows of eight pogo pins. In the Wellesley system, the electron signal is amplified in the gas using a GEM, and the SAQ front-end is contained within the vacuum vessel. Digital signals are sent through the vessel wall to the back-end. In the UTA system, no GEM is used (the final field shaping ring is grounded via a 50\,\Mohm{} resistor), the front-end board is outside of the vessel, and 16 copper traces on inner layers of the anode PCB serve as electrical vacuum feedthroughs for the analog signals.}
 \end{figure}

The anode plane is segmented into a series of concentric annuli, as shown in figure~\ref{fig:bothSystems}, and the spatial distribution of charge is reconstructed from the difference in the rate of resets on each annulus as measured with the Q-Pix prototype readout. 
The TPC, ionization source, and anode plane, as well as the SAQ prototype readout are shown schematically in figure~\ref{fig:bothSystems}, and described in more detail below.

\subsection{Gas TPC and ionization source}

The TPC is defined by a solid metal cathode, a series of field shaping rings with a resistor chain to establish a uniform drift field, and the segmented anode. The concentric anode annuli are patterned onto a custom printed circuit board (PCB). 
While this anode segmentation is not optimal for generic track reconstruction, it was chosen for this particular application because of the limited number of available readout channels (16), and the known and fixed location of the ionization source (approximately above the center of the anode plane). 
Internal PCB traces are routed from each ring to 16 exposed conductive pads. The SAQ readout connects to those pads via pogo pins. 

The TPC was housed inside of a vacuum vessel, which was evacuated and back-filled with the target gas (P-10) to the desired operating pressure. Outside of the vacuum vessel, a pulsed xenon flashlamp (XeF) served as the UV light source, generating drift electrons via the photoelectric effect. Light from the XeF was coupled into a fiber which entered the vacuum vessel via a fiber feedthrough (Accuglass 105201). All fiber components were multimode, 600\,$\mu$m diameter fused silica, with negligible attenuation at the relevant wavelengths. Inside the chamber, the fiber was threaded through the field cage with its bare end abutting a gold foil affixed to the cathode. 

Two variations of this system were constructed, one at Wellesley College and the other at the University of Texas, Arlington (UTA). Both systems employed the same segmented anode plane, and front-end and back-end electronics, but featured several distinct design elements, as outlined below.

\subsubsection{Configuration specific to the Wellesley system}
In the Wellesley system, the entire anode PCB, along with the front-end electronics, were housed within the vacuum vessel. The field cage consisted of four aluminum field shaping electrodes, each separated by 6.4\,mm insulating spacers and connected by $R=5\,\Mohm{}$ resistors, for a total drift length of 5.2\,cm. A standard CERN thin Gas Electon Multiplier (GEM) \cite{cernGEM} provided modest amplification (${\sim}10^2$) of the electron signal in the gas. The GEM is made of copper-clad kapton (50\,$\mu$m thick) with $5\times 5$\,cm$^2$ active area and bi-conical holes (50/70\,$\mu$m inner/outer diameter) on a triangular pattern with 140\,$\mu$m pitch. A high-voltage divider box external to the vacuum vessel (depicted in figure~\ref{fig:bothSystems}) maintained a fixed ratio of 30:1 between the GEM amplification and collection fields (the bottom of the GEM is 0.58\,mm above the anode). Two power supplies ($-$HHV and $-$HV in the figure) allow for independent adjustment of the drift field and GEM amplification fields. Large series resistors (100\,\Mohm{}) protect the GEM by limiting the discharge current in the event of a spark. 
A Hamamatsu L11316-11, average power of 5\,W, with user adjustable repetition rate, typically 1--10\,Hz was used as the UV light source.
Digital reset signals from the front-end are coupled to the back-end via a multi-pin electrical vacuum feedthrough. 

\subsubsection{Configuration specific to the UTA system}
The UTA system employed a cylindrical stainless steel vacuum vessel to enclose the TPC, with an o-ring seal between the vessel and the anode PCB. All readout electronics were located outside of the vacuum vessel, with internal PCB traces serving as electrical feedthroughs to the SAQ front-end board. The field cage consisted of nine rings (2.29\,mm thick) separated by insulating spacers (8.5\,mm thick), with $R=50\,\Mohm{}$ (see figure~\ref{fig:bothSystems}). The final field cage ring was connected to ground via a 50\,\Mohm{} resistor, and the total drift length was 10\,cm. No GEM (or high-voltage divider box) was used. 
 The UV light source was a Hamamatsu L13651-11, with average power of 2\,W, and operated at a repetition rate of 100\,Hz.

\subsection{SAQ readout}
\begin{figure}[t]
\centering
\includegraphics[width=0.9\textwidth]{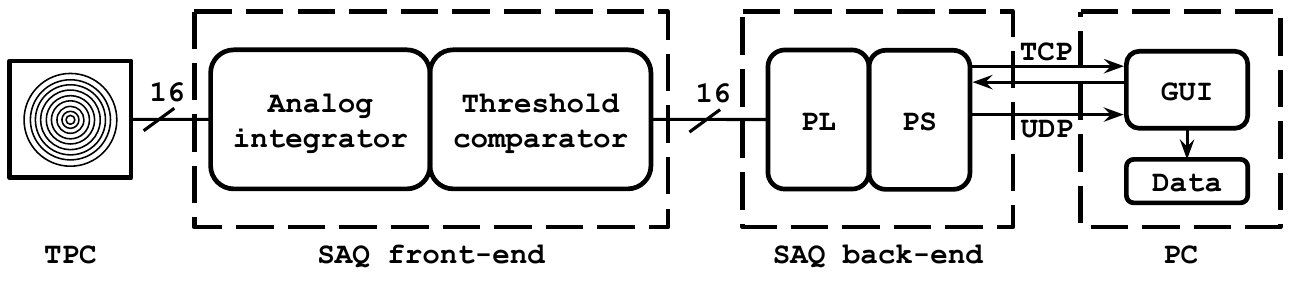}
\caption{\label{fig:SystemLayout}System block diagram of the SAQ front-end and back-end. Details of the TPC are provided in figure~\ref{fig:bothSystems}. Detector current is coupled to the front-end, which produces digital pulses each time a specified amount of charge has accumulated. Those digital reset pulses are processed by the back-end, which is implemented with both programmable logic (PL) and a processing system (PS) in a Zynq-7000 system on chip on a Digilent Zybo-Z7-20 board. The back-end communicates with a PC, sending detector reset data via UDP, and sending/receiving configuration parameters via TCP.}
\end{figure}

The ionization charge in the TPC was measured with our SAQ prototype. As shown in the block diagram of the readout system (figure~\ref{fig:SystemLayout}), the 16 anode electrodes are connected to the SAQ front-end. Each front-end channel integrates signal current from its corresponding anode electrode and generates a digital pulse upon reaching a preset threshold. The SAQ back-end then timestamps those reset pulses and transmits them to a computer for offline processing.

\begin{figure}[t]
 \centering 
 \includegraphics[width=0.9\textwidth]{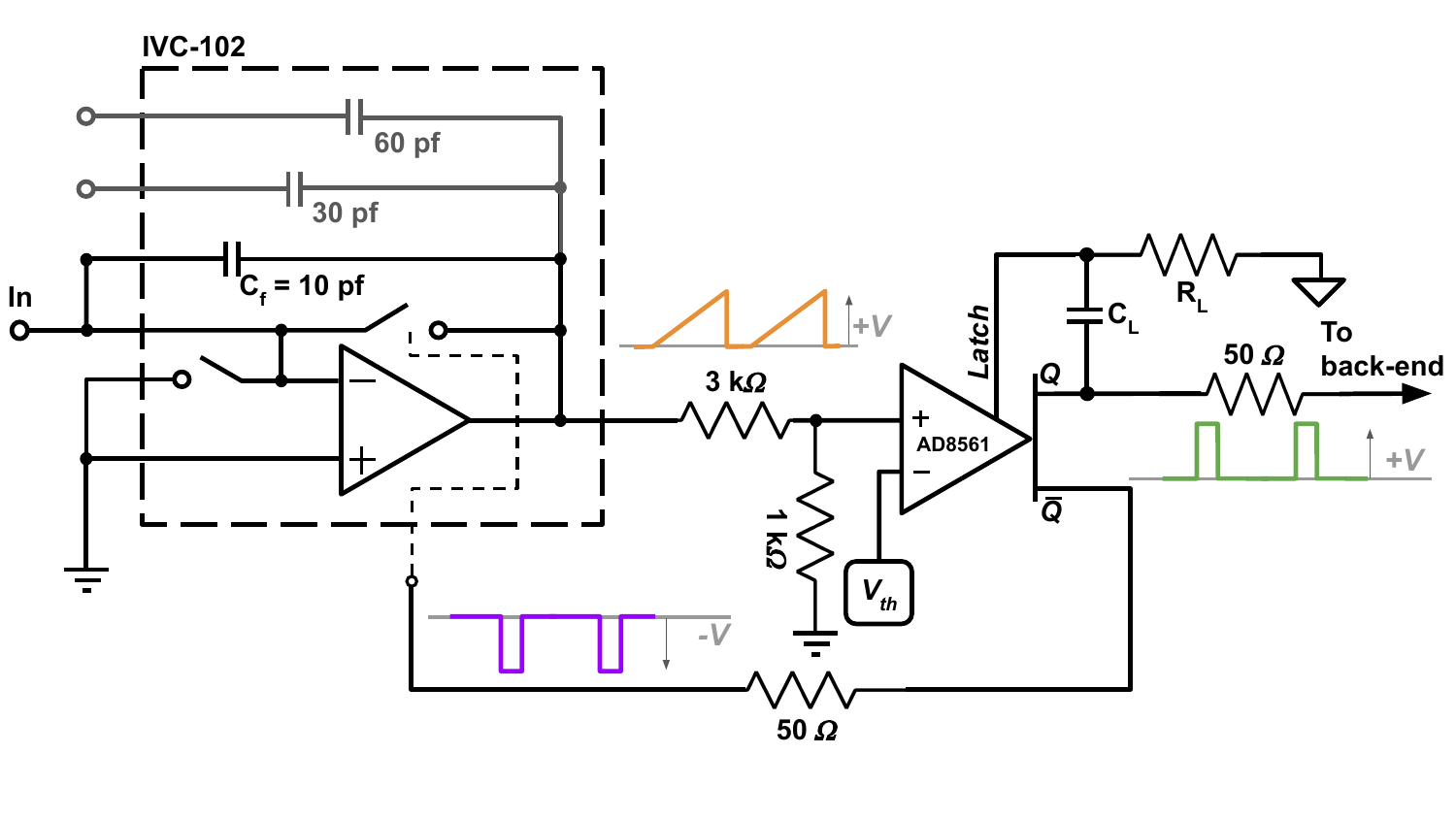}
 \caption{\label{fig:saqCircuit} Schematic of a single front-end channel (the SAQ front-end consists of 16 of these). Signal charge from the TPC is coupled to the summing junction of the TI IVC102 integrator (left). We operate with $C_f=10$\,pF, though values up to 100\,pF are available (shown in gray). When the integrator output exceeds the integration threshold (set by \Vth{} and the 4:1 divider), the comparator (AD8561) sends an active-low reset pulse $\bar{Q}$ to the integrator, and an active-high logic pulse $Q$ to the back-end to be timestamped and recorded. The width of the reset/logic pulse is set to 10\,$\mu$s via $R_L$ and $C_L$ at the Latch pin of the comparator. The example waveform shown at the integrator output (orange, linear ramp) corresponds to a constant input current.}
 \end{figure}
 
\subsubsection{SAQ analog front-end}\label{sec:saqFrontEnd}

As shown in figure~\ref{fig:saqCircuit}, each of the 16 front-end channels consists of a precision switched integrator transimpedance amplifier (Texas Instruments IVC102) followed by a latched comparator (Analog Devices AD8561). The IVC102 has a selectable feedback capacitance $C_f$ of 10 to 
100\,pF, which we operate at the most sensitive setting of 10\,pF (smallest $\Delta Q$).
The integrator output drives a comparator whose other input is maintained at a user-adjustable threshold voltage \Vth{}. A resistive divider between the integrator and comparator reduces the integrator output by a factor of four. When the integrator output (divided by four) exceeds \Vth{}, the comparator generates a fixed-width pulse $Q$ and its complement $\bar{Q}$. The active low $\bar{Q}$ is fed back to the IVC102 to reset the integrator, while the active high $Q$ alerts the back-end that a reset has occurred (see section~\ref{sec:saqBackEnd}). The IVC102 specifications indicate that the reset duration $t_R$ must be at least 10\,$\mu$s to completely discharge $C_f$. We achieve this with the latch function of the comparator, where an external resistor and capacitor ($R_L$ and $C_L$ in figure~\ref{fig:saqCircuit}) determine the pulse width. We operate with $t_R=10\,\mu$s to minimize the loss of ionization signal from the TPC (dead time). During our measurements, the dead time was negligible, accounting for less than 0.1\% of the time between resets.\footnote{For the Q-Pix ASIC, a zero-dead-time replenishment scheme, rather than a reset scheme, is used.}
A typical reset threshold voltage during operation was 1\,V at the integrator output, corresponding to an integrated charge of 10\,pC (${\sim} 6\times 10^7$ electrons). In our system, multiple XeF pulses were required to accumulate enough charge for a single reset, resulting in an averaged measurement of the charge cloud geometry over many pulses. This approach sacrifices information about the extent of the charge distribution in the drift direction. Lowering the reset threshold to a level where a single XeF pulse produces multiple resets would enable a measurement of the longitudinal diffusion by providing information about the spatial extent of the charge cloud in the drift direction.

\subsubsection{SAQ digital back-end}\label{sec:saqBackEnd}

The SAQ digital back-end is implemented in an FPGA that receives and timestamps reset signals ($Q$) from the 16 front-end channels, and transmits that data via UDP to a host computer. A python-based GUI on the host computer displays real-time data and records the data to disk for offline analysis. In addition, the GUI provides an interface to configure the FPGA via TCP, allowing the user to select experimental parameters such as which channels are active.

The FPGA is a Zynq-7000 System on Chip (SoC) on a Zybo-Z7-20 Digilent evaluation board, operating with a 30.3\,MHz clock and using an ARM Cortex-A9 to communicate with the Programmable Logic (PL). The PL records a timestamp from a reset signal as a 32-bit counter on the next clock cycle after the reset. The timestamp and the channel mask (indicating which channels reset) are accumulated in a First-In-First-Out (FIFO) buffer. If at least one of the 16 digital input channels transitions from low to high, and the FIFO register is enabled, then data are written to the FIFO. The input from all 16 channels are recorded at the time of the trigger, regardless of which channel caused the trigger. 
Detecting the rising edge of the reset pulse ensures that a single reset cannot generate multiple triggers -- in order for a new reset to be recorded on the same channel, the reset pulse must first be driven low for at least one clock cycle ($\geq$ 33 ns) to reset the latch at the input channel. 
A detailed description of the back-end firmware and embedded software is available in a recent PhD thesis~\cite{kevinThesis}.

\section{Transverse diffusion measurements}\label{sec:measurements}
We measured the transverse diffusion of electrons in a gas TPC filled with P-10 at a range of pressures and drift electric fields. As shown in figure~\ref{fig:dbymu}, we expect the transverse diffusion \sigmaDrift{} to decrease with increasing pressure at fixed field, but remain approximately constant with field at fixed pressure. Spatial variations of the charge in the TPC manifest as channel-to-channel differences in the amount of deposited charge on each concentric annular electrode, producing different reset rates in the SAQ readout system. We describe our channel-to-channel calibration procedure used to determine the expected charge per reset. We also describe the expected signal in the likely case that the primary photoelectron cloud is neither symmetric nor aligned with the center of the anode. We then present charge distribution profiles measured over a range of gas pressures and drift fields, and compare these measurements to a detector simulation to determine the diffusion.

\subsection{SAQ system calibration}\label{sec:calibration}

\begin{figure}[t]
    \centering
    \includegraphics[width=0.95\textwidth]{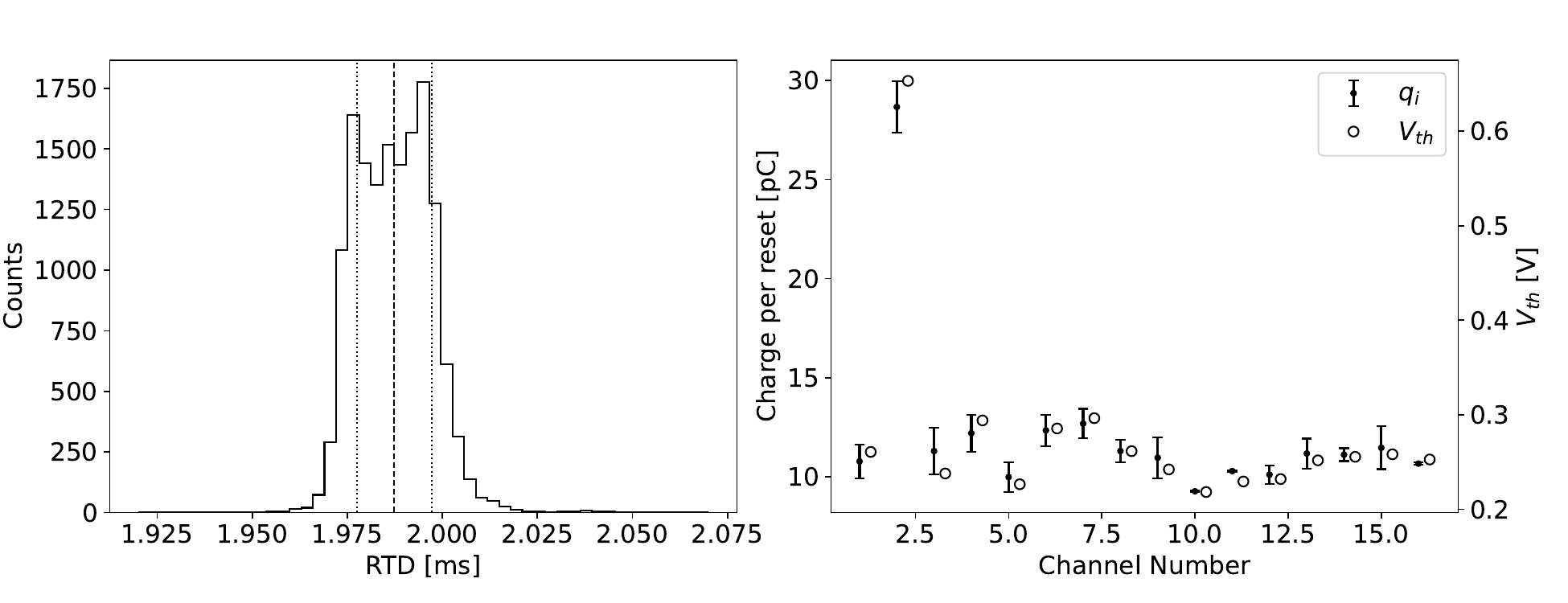}
    \caption{\label{fig:qPerReset}(Left) Reset Time Differences (RTD) measured by one SAQ front-end channel for a constant input current.
    The vertical dashed line and dotted lines indicates the mean and standard deviation of the distribution. The $0.5\%$ resolution (standard deviation divided by mean) arises primarily from environmental noise coupled to the front-end electronics. (Right) Calibrated charge per reset $q_i$ for all 16 SAQ channels (filled circles) and the measured reset threshold voltage (open circles). For visual clarity, threshold data points are shifted slightly to the right to avoid overlap with the corresponding charge per reset data. The threshold for channel 2 was set higher than the rest by a factor of ${\sim}3$. }
\end{figure}

An accurate diffusion measurement requires the calibration of channel-to-channel variations in the system response. Given a spatially uniform charge distribution in the detector,  variations in the reset rate across channels can be attributed to two main factors: (1) differences in the reset thresholds, and (2) annular area differences for each concentric anode ring. We calibrate the former by recording the distribution of reset time differences (RTDs) for a known constant applied current \Iconst{} at the input to the front-end for each channel. We find that the variations in the mean RTD across channels are indeed explained by differences in the user-defined comparator thresholds \Vth{}. This is easily understood since the accumulated charge in time $\Delta t$ is $q = \int \Iconst dt = \Iconst \Delta t$, and a channel will execute a reset after a time
\begin{equation}
    \Delta t_{\mbox{\tiny reset}} = \mbox{RTD} = \frac{C_f (4\Vth)}{\Iconst},
\end{equation}
where the factor of 4 in the numerator accounts for the resistive divider between the integrator output and the comparator. So $q_i$, the charge required for a reset on channel $i$ is proportional to the threshold voltage \Vth{} on that channel.
Uncertainties in the measured values of $q_i$ are given by the standard deviations of the RTD distributions. We repeat this calibration for a range of input currents, and find consistent results 
across two orders of magnitude for \Iconst{} for the duration of the experiment. This result is summarized in figure~\ref{fig:qPerReset}, which shows the RTD distribution for one front-end channel. The figure also shows the measured $q_i$ values and uncertainties for all 16 front-end channels, in comparison with a measurement of \Vth{} for the same channels. Their close correlation indicates that the variations in the charge per reset for each channel is well-understood, and can be easily compensated for in the analysis.

The area of each anode segment increases with radius to compensate for the expected reduction in signal charge at large radii. When reconstructing the transverse charge distribution we normalize the number of resets (and therefore the total accumulated charge) to the area of each anode annulus. Combining this with the calibrated charge per reset, we then report $\QofR{}$, the reconstructed charge per area as a function of radius in the anode plane. We then determine the transverse diffusion by comparing \QofR{} to a simulation of the compact charge cloud transport from the cathode to the anode in the TPC.

\subsection{Fiber orientation and measured charge distribution}\label{sec:diffusionModel}
\begin{figure}[t]
    \centering
    \includegraphics[width=0.95\textwidth]{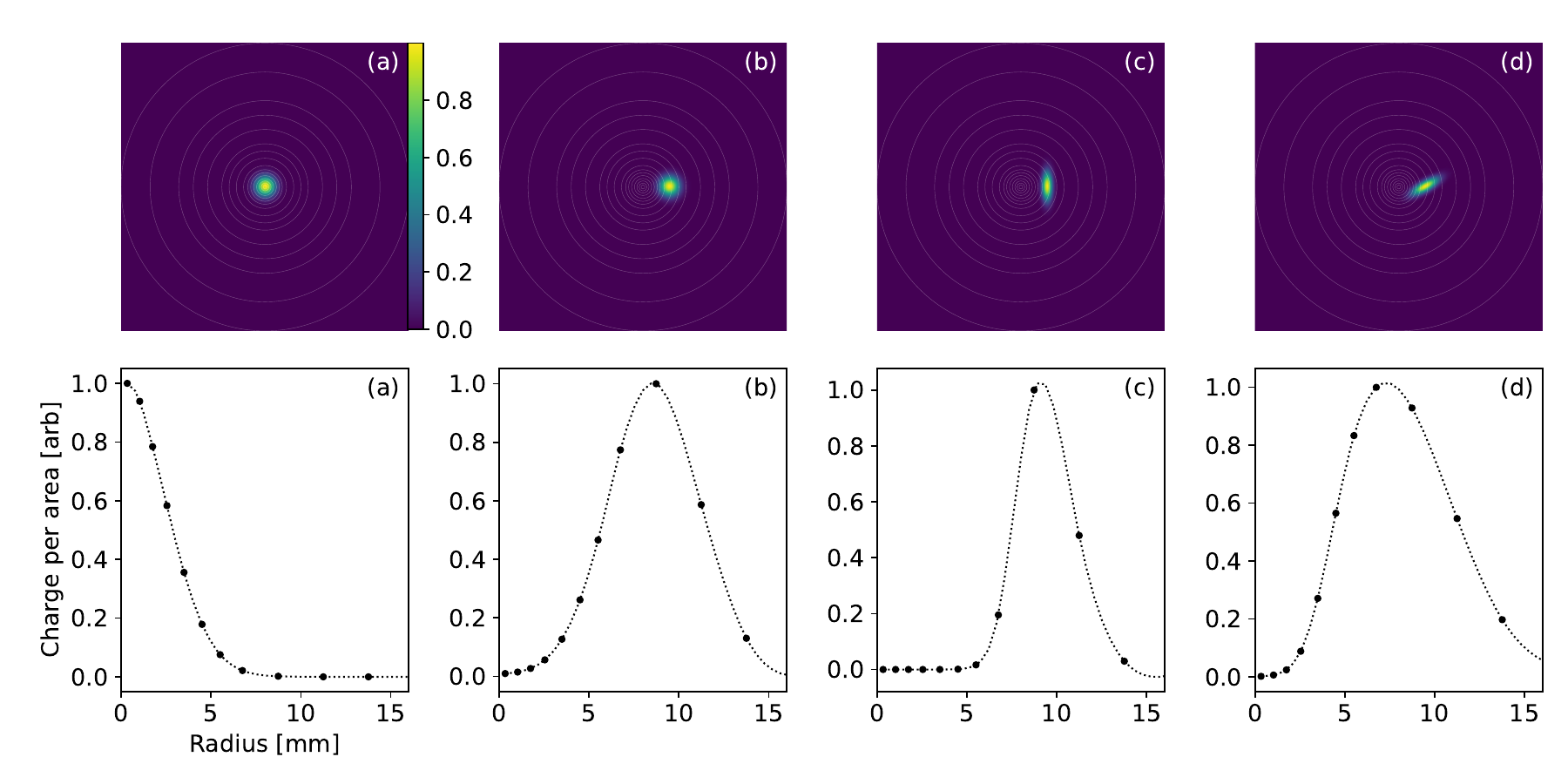}
    \caption{(Top row) Simulated electron cloud distributions at the photocathode for four different fiber orientations. Thin white lines indicate boundaries between the 16 concentric anode electrodes. The radius of the outermost circle is 50\,mm. (Bottom row) Simulated distributions of charge per area as a function of radius, \QofR{}, corresponding to the images in the top row. Each point corresponds to the integrated electron signal in each circular annulus, normalized by the area of each annulus.
    (a) Fiber axis is normal to the photocathode and centered on the anode axis. The illumination pattern is a symmetric 2D gaussian, and \QofR{} is a gaussian centered at the origin.
    (b) Same as (a) but with the fiber axis offset by a distance $\mu_r=9$\,mm from the center of the anode.
    (c) Fiber offset from the origin, as in (b), but tilted so that the angle of the fiber axis relative to the photocathode normal is $\theta=70^\circ$ (with $\phi=0$).
    (d) Same as (c) but fiber is rotated in the azimuthal direction by $\phi=60^\circ$. 
    }
    \label{fig:diffExp}
\end{figure}

As explained in section~\ref{sec:saq}, a point-like charge cloud produced directly above the center of the anode will produce a gaussian \QofR{} profile centered at $r=0$ whose width grows with the amount of transverse diffusion (see equation~\ref{eq:diffgaus} and figure~\ref{fig:diffExp}(a)). In our setups, 
the primary charge cloud may not be perfectly aligned with the center of the anode, nor will it start out circularly symmetric in the cathode plane because of the orientation of the fiber tip relative to the gold foil. 

We model the dependence of \QofR{} on fiber orientation by assuming that when the fiber is normal to the gold foil the UV light emerging from the fiber produces a symmetric 2D gaussian illumination pattern with width \sigmaInitial{}. We further assume that the spatial distribution of photoelectrons at creation matches the UV light distribution.
If the fiber axis were normal to the cathode, but off-center by a distance $\mu_r$, then the photoelectron distribution would be a symmetric 2D gaussian offset from the center of the anode. As shown in figure~\ref{fig:diffExp}(b), the associated \QofR{} distribution is non-gaussian (nor is it symmetric) due to the annular segmentation of the anode plane. 
We must also account for the fiber's angular alignment relative to the gold foil. The fiber enters our field cage through a gap in the upper rings such that the angle between the photocathode normal and the fiber axis is $\theta \approx 70^\circ$ (see figure~\ref{fig:bothSystems}). 
The resulting illumination pattern (and, by assumption, the resulting charge distribution) is an asymmetric 2D gaussian, as seen in figure~\ref{fig:diffExp}(c).
A second angle $\phi$, azimuthal in the cathode plane, is needed to fully specify the fiber axis. We define $\phi=0$ to align with the vertical direction in the images of figure~\ref{fig:diffExp} so that $\phi=0^\circ$ in (c). Plot (d) in that figure shows how \QofR{} changes with $\phi$, with all other parameters held constant relative to (c).

We independently measure the fiber orientation to constrain these nuisance parameters ($\mu_r$, \sigmaInitial{}, $\theta$ and $\phi$) in the diffusion analysis. Differences in the TPC design between UTA and Wellesley require different approaches to determining these constraints. Specifically, the Wellesley field cage is attached to the anode plane, so direct imaging of the fiber illumination pattern on the gold photocathode was not possible. Instead, measurements of the fiber orientation angles were made, leading to increased uncertainties in the measured diffusion. The UTA field cage, on the other hand, is suspended from the vacuum vessel lid, and so direct imaging of the UV illumination pattern on the gold photocathode was possible. 
These nuisance parameters are stable over the course of a series of data sets (e.g. for a range of pressure values at a fixed electric field), and are therefore fixed, while the transverse diffusion is allowed to vary.

\subsection{Transverse diffusion measurements}
Measurements of electron diffusion were made over a range of drift field and gas pressures. Here, we report the results of two experiments: first, a pressure scan in which the drift field was held fixed but the gas pressure was varied, and second, a drift field scan in which the pressure was held constant but the drift field varied.

\begin{figure}
    \centering
    \includegraphics[width=0.98\textwidth]{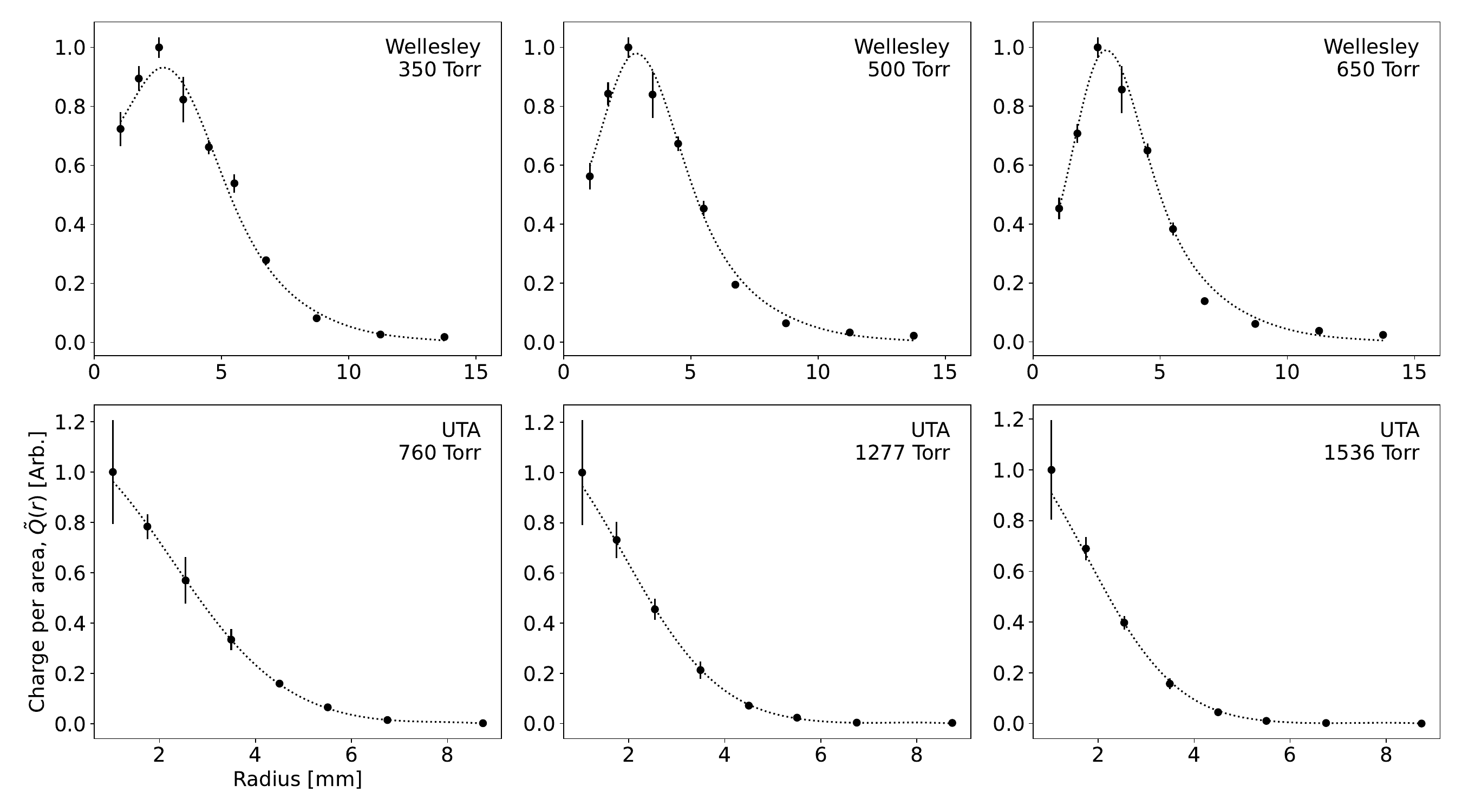}
    \caption{Representative \QofR{} measurements from the pressure scan with the Wellesley apparatus (top row) and the UTA apparatus (bottom row), with pressures indicated in the plots. Data are taken at a fixed drift field (400\,V/cm for Wellesley and 500\,V/cm for UTA). Black points show the measured charge per area for each anode segment. Uncertainties are determined from the calibration shown in figure~\ref{fig:qPerReset}. As expected, the distributions narrow with increasing pressure, indicative of decreasing transverse diffusion \sigmaDrift{}. Data are normalized to unit amplitude to account for variations in the duration of each dataset as well as signal strength differences between the Wellesley and UTA systems. Model fits, shown as dotted curves, use a single set of fiber orientation parameters, but allow for different amounts of diffusion \sigmaDrift{} at each pressure. A comparison with figure~\ref{fig:diffExp} indicates that the fiber in the Wellesley setup was offset from the center of the anode by several millimeters, while the UTA one was not.}
    \label{fig:pScan}
\end{figure}

\begin{figure}
    \centering
    \includegraphics[width=0.98\textwidth]{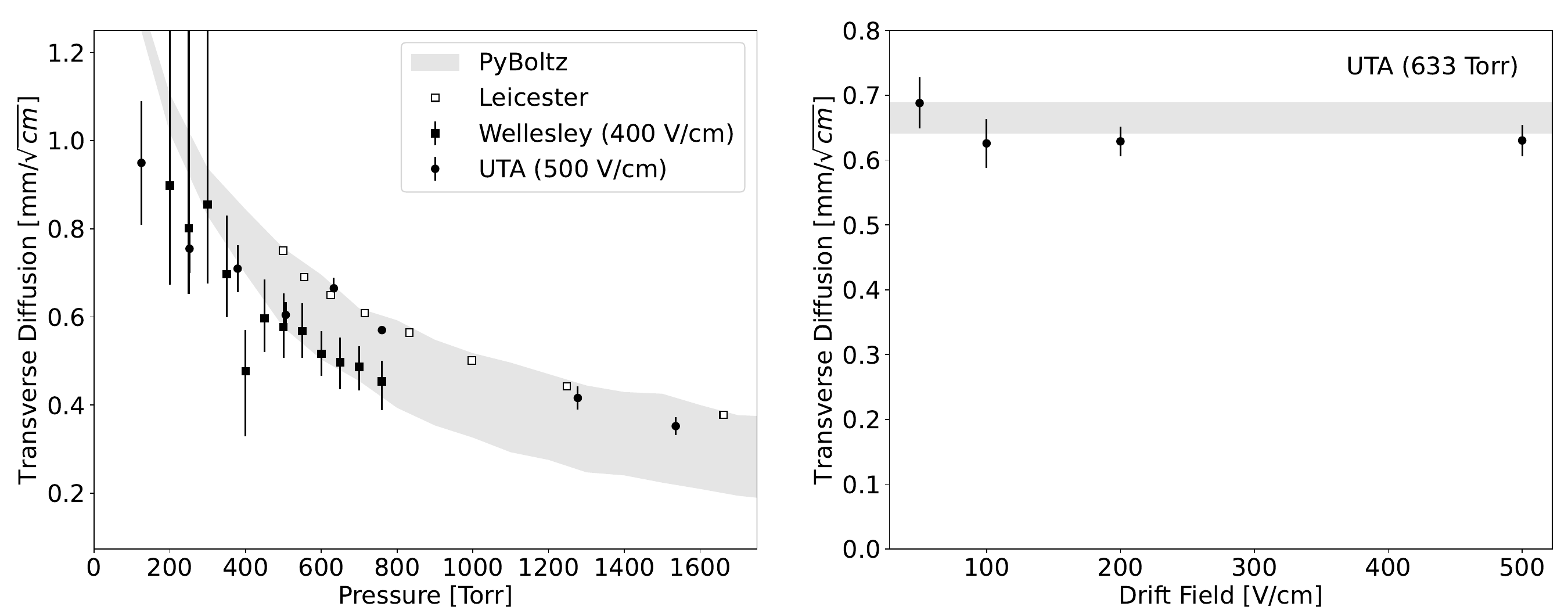}
    \caption{(Left) Transverse diffusion measurements, $\sqrt{\sigmaDrift^2+\sigmaOther^2}$, from a scan of gas pressure at fixed drift field ($E=400$\,V/cm for Wellesley, white circles; 500\,V/cm for UTA, black circles). The gray band represents the transverse diffusion predicted by the PyBoltz simulation at a drift field of 500\,V/cm, and is based on the simulation of $D_T/\mu$ shown in figure~\ref{fig:dbymu}.
    The upper edge of the band corresponds to pure P-10, while the lower edge of the band includes the effect of water vapor contamination (99\% P-10, 1\% H$_2$O). Independent measurements from the University of Leicester of transverse diffusion in P-10 using the Townsend method in a drift chamber are shown as white squares~\cite{leicester1984}.
    (Right) Measured transverse diffusion as a function of drift field, for a fixed pressure (633\,Torr, UTA apparatus). The gray band indicates the diffusion (with uncertainty) measured with the same apparatus and pressure, with a drift field of 500\,V/cm during the pressure scan shown in the left plot.
    }
    \label{fig:diffDataVsPyBoltz}
\end{figure}

For the pressure scan, the vacuum vessel was evacuated and then back-filled with P-10 to the desired pressure. The xenon flashlamp was then enabled, with a fixed pulse repetition rate and recorded resets on each anode element with the SAQ. This process was then repeated at each pressure and 
 the associated charge per area distributions \QofR{} were constructed using the calibrations described in section \ref{sec:calibration}. A sample of the resulting charge per area profiles \QofR{} is shown in figure~\ref{fig:pScan}. As the pressure increases, \QofR{} clearly narrows indicating a smaller transverse diffusion. This trend is quantified by fitting models of \QofR{} (as in figure~\ref{fig:diffExp}) to extract the transverse diffusion. The fits (also shown in figure~\ref{fig:pScan}) are constrained to use the same fiber orientation parameters for all pressures in the scan, allowing only the diffusion to vary.
 The resulting measurement of transverse diffusion as a function of pressure is shown in the left plot of figure~\ref{fig:diffDataVsPyBoltz}. 

 Also shown in figure~\ref{fig:diffDataVsPyBoltz} (left) are expectations based on simulation and prior measurements. The simulated transverse diffusion is obtained by combining the PyBoltz calculation of $D_T/\mu$ (figure~\ref{fig:dbymu}) with equation~\ref{eq:sigmaT}, which relates $D_T/\mu$ and \sigmaDrift{}. To explore the effect of water vapor contamination of the target gas, the PyBoltz simulation was repeated for a gas mixture containing 99\% P-10 and 1\% H$_2$O. The prior measurements of diffusion in this range of drift fields and gas pressures come from a 1984 PhD thesis from the University of Leicester, and were made using the Townsend method in a drift chamber~\cite{leicester1984}. Those measurements agree well with the PyBoltz prediction for pure P-10.

To validate the pressure scan results, the transverse diffusion was also measured over a range of drift fields at fixed pressure.
Based on the linear dependence of $D_T/\mu$ with $E$ for P-10 as shown in figure~\ref{fig:dbymu}, \sigmaDrift{} should be independent of $E$. As expected, our measurements of diffusion show no dependence on drift field. As an example, figure~\ref{fig:diffDataVsPyBoltz} (right) shows results from the UTA system at constant pressure (633\,Torr) and for a range of drift field strengths spanning more than an order of magnitude (50 to 500\,V/cm).

\subsection{Discussion}

We have demonstrated the successful operation of a multichannel Q-Pix prototype in which we reconstructed information about the topology of ionization in a gas TPC from the timestamps of resets. Our measurements of the transverse diffusion of electrons follow expected trends with pressure and drift field, and are consistent with expectations based on simulation and measurements from another group using a different technique.

Ideally, our measurements would be made in pure P-10, but small leaks in the vacuum vessel and outgassing from internal components introduce impurities like water vapor to the gas. Although efforts were taken to mitigate this effect, it is reasonable to expect modest gas contamination. As described in section~\ref{sec:diff}, molecular contaminants like water with closely spaced internal energy levels can significantly lower the characteristic electron energy, which suppresses transverse diffusion. Although we were not able to directly measure the contamination levels in our systems, we explored this effect through PyBoltz simulations of the diffusion in both pure P-10 and a 99\%:1\% mixture of P-10 and water vapor. The Wellesley measurements, and the low-pressure UTA ones lie below the expectations for pure P-10, consistent with the presence of impurities (see figure~\ref{fig:diffDataVsPyBoltz}, left).

A competing effect is that the measured diffusion reported in figure~\ref{fig:diffDataVsPyBoltz} depends on the quadrature sum of the diffusion due to drift \sigmaDrift{} and other contributions to the measured width of the charge cloud \sigmaOther{} (see equation~\ref{eq:sigmaTot}). Here, \sigmaOther{} is specific to the apparatus, and represents the width that would be measured for a point-like ionization cloud and zero drift length. While it is impractical to achieve zero drift length, one could measure \sigmaOther{} by, for example, generating primary ionization at different initial heights $z$ in the TPC and extrapolating the measured $\sigmaTot^2(z)$ to zero drift length (for an example of this, see Ref.~\cite{Battat2014}). In our system, we generated photoelectrons at the cathode and are not able to adjust the drift length (short of rebuilding the TPC), so we did not disentangle the transverse diffusion \sigmaDrift{} from $\sigmaOther{}$. This tends to overestimate \sigmaDrift{}, but is a limitation that arises from the way we generate our ionization cloud, and not from the Q-Pix technique itself.

One contributor to \sigmaOther{} in the Wellesley apparatus is diffusion in the collection region between the bottom of the GEM and the anode plane. However, we argue that the collection diffusion is negligible compared to the diffusion in the drift region. As shown in figure~\ref{fig:dbymu} and the rightmost plot in \ref{fig:diffDataVsPyBoltz}, transverse diffusion in P-10 does not vary with electric field. Equation~\ref{eq:sigmaT} indicates that the ratio of transverse diffusion due to drift to the diffusion in the collection field will therefore scale as the square root of the ratio of the lengths of those two regions (52\,mm and 0.5\,mm), and the diffusion in the collection region will be an order of magnitude smaller than in the drift region.

\section{Conclusion}\label{sec:conclusion}
Using a 16-channel Q-Pix prototype constructed from commercial-off-the-shelf components, we have measured the transverse diffusion of electrons in P-10 gas in a TPC as a function of gas pressure and drift electric field. The results align with predictions using PyBoltz simulations, as well as previous diffusion measurements from the literature. This study demonstrates successful operation of a multi-channel Q-Pix analog front-end and FPGA back-end to reconstruct the topology of ionization in a TPC from the timestamps of reset pulses. Together with the related work of the demonstration of a high-fidelity reconstruction of an input current waveform with a COTS Q-Pix prototype front-end \cite{Miao:2023ivo}, these results are a promising step toward a low-power and low-threshold ASIC-based pixelated TPC readout for 3D track imaging and calorimetry. Such a system would have wide application in neutrino detection and other  areas of particle and nuclear physics.

\acknowledgments
We thank Jim MacArthur, director of the Electronics Instrument Design Laboratory at Harvard University for consulting on the SAQ electronics design. We also thank the following members of JBRB's research group who contributed to the construction and operation of the Wellesley detector: Eunice Beato, Debra Lacey, Diana Lopez, Natalie McGee and Julia Sherman. This work was supported by the Department of Energy through a FAIR Award DE-SC0024323, Office of High Energy Physics Award No.~DE-0000253485 and No.~DE-SC0020065, the Gordon and Betty Moore Foundation through Grant GBMF11565 and grant DOI https://doi.org/10.37807/GBMF11565, and  
the Science and Technology Facilities Council (STFC), part of the United Kingdom Research and Innovation and the UK Research and Innovation (UKRI) Ernest Rutherford Fellowship. 
S. Kubota is supported by the Masason Foundation and the Ezoe Memorial Recruit Foundation.

\bibliographystyle{JHEP}
\bibliography{biblio}

\providecommand{\href}[2]{#2}\begingroup\raggedright\begin{thebibliography}{10}

\bibitem{Nygren:2018rbl}
D.~Nygren and Y.~Mei, \emph{{Q-Pix: Pixel-scale Signal Capture for Kiloton Liquid Argon TPC Detectors: Time-to-Charge Waveform Capture, Local Clocks, Dynamic Networks}},  \href{https://arxiv.org/abs/1809.10213}{{\ttfamily 1809.10213}}.

\bibitem{ARNEODO199795}
F.~Arneodo et~al., \emph{The liquid argon {TPC} for the {ICARUS} experiment}, \href{https://doi.org/https://doi.org/10.1016/S0920-5632(97)00098-4}{\emph{Nuclear Physics B - Proceedings Supplements} {\bfseries 54} (1997) 95}.

\bibitem{Acciarri_2017}
R.~Acciarri et~al., \emph{Design and construction of the {MicroBooNE} detector}, \href{https://doi.org/10.1088/1748-0221/12/02/p02017}{\emph{Journal of Instrumentation} {\bfseries 12} (2017) P02017}.

\bibitem{Acciarri_2020}
R.~Acciarri et~al., \emph{The liquid argon in a testbeam ({LArIAT}) experiment}, \href{https://doi.org/10.1088/1748-0221/15/04/p04026}{\emph{Journal of Instrumentation} {\bfseries 15} (2020) P04026}.

\bibitem{Anderson:2012vc}
C.~Anderson, M.~Antonello, B.~Baller, T.~Bolton, C.~Bromberg, F.~Cavanna et~al., \emph{{The ArgoNeuT Detector in the NuMI Low-Energy beam line at Fermilab}}, \href{https://doi.org/10.1088/1748-0221/7/10/P10019}{\emph{JINST} {\bfseries 7} (2012) P10019} [\href{https://arxiv.org/abs/1205.6747}{{\ttfamily 1205.6747}}].

\bibitem{Zhang_2011}
H.~Zhang et~al., \emph{The {ATLAS} liquid argon calorimeter: Overview and performance}, \href{https://doi.org/10.1088/1742-6596/293/1/012044}{\emph{Journal of Physics: Conference Series} {\bfseries 293} (2011) 012044}.

\bibitem{WILLIS1974221}
W.~Willis and V.~Radeka, \emph{Liquid-argon ionization chambers as total-absorption detectors}, \href{https://doi.org/https://doi.org/10.1016/0029-554X(74)90039-1}{\emph{Nuclear Instruments and Methods} {\bfseries 120} (1974) 221}.

\bibitem{Abi_2020}
B.~Abi et~al., \emph{Volume {IV}. the {DUNE} far detector single-phase technology}, \href{https://doi.org/10.1088/1748-0221/15/08/t08010}{\emph{Journal of Instrumentation} {\bfseries 15} (2020) T08010}.

\bibitem{DUNE:2020cqd}
{\scshape DUNE} collaboration, \emph{{First results on ProtoDUNE-SP liquid argon time projection chamber performance from a beam test at the CERN Neutrino Platform}}, \href{https://doi.org/10.1088/1748-0221/15/12/P12004}{\emph{JINST} {\bfseries 15} (2020) P12004} [\href{https://arxiv.org/abs/2007.06722}{{\ttfamily 2007.06722}}].

\bibitem{Hernandez:2020fpm}
A.S.~Hern\'andez, D.~Gonz\'alez-D\'\i{}az, P.~Villanueva, C.~Azevedo and M.~Seoane, \emph{{A new imaging technology based on Compton X-ray scattering}},  \href{https://arxiv.org/abs/2006.01504}{{\ttfamily 2006.01504}}.

\bibitem{GRIGNON2007142}
C.~Grignon, J.~Barbet, M.~Bardiès, T.~Carlier, J.~Chatal, O.~Couturier et~al., \emph{Nuclear medical imaging using $\beta + \gamma$ coincidences from $^{44}${Sc} radio-nuclide with liquid xenon as detection medium}, \href{https://doi.org/https://doi.org/10.1016/j.nima.2006.10.048}{\emph{Nuclear Instruments and Methods in Physics Research Section A: Accelerators, Spectrometers, Detectors and Associated Equipment} {\bfseries 571} (2007) 142}.

\bibitem{CHEPEL1997427}
V.~Chepel, M.~Lopes, A.~Kuchenkov, R.~{Ferreira Marques} and A.~Policarpo, \emph{Performance study of liquid xenon detector for pet}, \href{https://doi.org/https://doi.org/10.1016/S0168-9002(97)00196-4}{\emph{Nuclear Instruments and Methods in Physics Research Section A: Accelerators, Spectrometers, Detectors and Associated Equipment} {\bfseries 392} (1997) 427}.

\bibitem{Acciarri_2010}
R.~Acciarri et~al., \emph{The {WArP} experiment}, \href{https://doi.org/10.1088/1742-6596/203/1/012006}{\emph{Journal of Physics: Conference Series} {\bfseries 203} (2010) 012006}.

\bibitem{AMAUDRUZ20191}
P.-A.~Amaudruz et~al., \emph{Design and construction of the {DEAP-3600} dark matter detector}, \href{https://doi.org/https://doi.org/10.1016/j.astropartphys.2018.09.006}{\emph{Astroparticle Physics} {\bfseries 108} (2019) 1}.

\bibitem{Collaboration_2009}
T.A.~Collaboration, V.~Boccone et~al., \emph{Development of wavelength shifter coated reflectors for the {ArDM} argon dark matter detector}, \href{https://doi.org/10.1088/1748-0221/4/06/p06001}{\emph{Journal of Instrumentation} {\bfseries 4} (2009) P06001}.

\bibitem{Aprile_2012}
E.~Aprile et~al., \emph{The {XENON}100 dark matter experiment}, \href{https://doi.org/10.1016/j.astropartphys.2012.01.003}{\emph{Astroparticle Physics} {\bfseries 35} (2012) 573}.

\bibitem{Akerib_2014}
D.~Akerib et~al., \emph{First results from the {LUX} dark matter experiment at the {Sanford Underground Research Facility}}, \href{https://doi.org/10.1103/physrevlett.112.091303}{\emph{Physical Review Letters} {\bfseries 112} (2014) }.

\bibitem{Anton_2019}
G.~Anton et~al., \emph{Search for neutrinoless double- $\beta$ decay with the complete {EXO}-200 dataset}, \href{https://doi.org/10.1103/physrevlett.123.161802}{\emph{Physical Review Letters} {\bfseries 123} (2019) }.

\bibitem{_lvarez_2013}
V.~{\'{A}}lvarez et~al., \emph{Operation and first results of the {NEXT}-{DEMO} prototype using a silicon photomultiplier tracking array}, \href{https://doi.org/10.1088/1748-0221/8/09/p09011}{\emph{Journal of Instrumentation} {\bfseries 8} (2013) P09011}.

\bibitem{Nygren_1974}
D.R.~Nygren, \emph{{The Time Projection Chamber: A New 4$\pi$ Detector for Charged Particles}}, .

\bibitem{antonello2015operation}
M.~Antonello, P.~Aprili, B.~Baibussinov, F.~Boffelli, A.~Bubak, E.~Calligarich et~al., \emph{{Operation and performance of the ICARUS-T600 cryogenic plant at Gran Sasso underground Laboratory}}, \href{https://doi.org/10.1088/1748-0221/10/12/P12004}{\emph{JINST} {\bfseries 10} (2015) P12004} [\href{https://arxiv.org/abs/1504.01556}{{\ttfamily 1504.01556}}].

\bibitem{acciarri2017design}
{\scshape MicroBooNE Collaboration} collaboration, \emph{{Design and Construction of the MicroBooNE Detector}}, \href{https://doi.org/10.1088/1748-0221/12/02/P02017}{\emph{JINST} {\bfseries 12} (2017) P02017} [\href{https://arxiv.org/abs/1612.05824}{{\ttfamily 1612.05824}}].

\bibitem{LArIAT:2019kzd}
{\scshape LArIAT Collaboration} collaboration, \emph{{The Liquid Argon In A Testbeam (LArIAT) Experiment}}, \href{https://doi.org/10.1088/1748-0221/15/04/P04026}{\emph{JINST} {\bfseries 15} (2020) P04026} [\href{https://arxiv.org/abs/1911.10379}{{\ttfamily 1911.10379}}].

\bibitem{Bian:2015qka}
J.~Bian, \emph{{The CAPTAIN Experiment}},  in \emph{{Meeting of the APS Division of Particles and Fields}}, 9, 2015 [\href{https://arxiv.org/abs/1509.07739}{{\ttfamily 1509.07739}}].

\bibitem{acciarri2016long}
{\scshape DUNE Collaboration} collaboration, \emph{{Long-Baseline Neutrino Facility (LBNF) and Deep Underground Neutrino Experiment (DUNE)}: {Conceptual Design Report, Volume 1: The LBNF and DUNE Projects}},  \href{https://arxiv.org/abs/1601.05471}{{\ttfamily 1601.05471}}.

\bibitem{Qian:2018qbv}
X.~Qian, C.~Zhang, B.~Viren and M.~Diwan, \emph{{Three-dimensional Imaging for Large LArTPCs}}, \href{https://doi.org/10.1088/1748-0221/13/05/P05032}{\emph{JINST} {\bfseries 13} (2018) P05032} [\href{https://arxiv.org/abs/1803.04850}{{\ttfamily 1803.04850}}].

\bibitem{MicroBooNE:2020vry}
{\scshape MicroBooNE Collaboration} collaboration, \emph{{Neutrino event selection in the MicroBooNE liquid argon time projection chamber using Wire-Cell 3D imaging, clustering, and charge-light matching}}, \href{https://doi.org/10.1088/1748-0221/16/06/P06043}{\emph{JINST} {\bfseries 16} (2021) P06043} [\href{https://arxiv.org/abs/2011.01375}{{\ttfamily 2011.01375}}].

\bibitem{MicroBooNE:2020jgj}
{\scshape MicroBooNE Collaboration} collaboration, \emph{{High-performance Generic Neutrino Detection in a LArTPC near the Earth's Surface with the MicroBooNE Detector}},  \href{https://arxiv.org/abs/2012.07928}{{\ttfamily 2012.07928}}.

\bibitem{MicroBooNE:2021zul}
{\scshape MicroBooNE Collaboration} collaboration, \emph{{Cosmic Ray Background Rejection with Wire-Cell LArTPC Event Reconstruction in the MicroBooNE Detector}}, \href{https://doi.org/10.1103/PhysRevApplied.15.064071}{\emph{Phys. Rev. Applied} {\bfseries 15} (2021) 064071} [\href{https://arxiv.org/abs/2101.05076}{{\ttfamily 2101.05076}}].

\bibitem{MicroBooNE:2021ojx}
{\scshape MicroBooNE Collaboration} collaboration, \emph{{Wire-Cell 3D Pattern Recognition Techniques for Neutrino Event Reconstruction in Large LArTPCs: Algorithm Description and Quantitative Evaluation with MicroBooNE Simulation}}, \href{https://doi.org/10.1088/1748-0221/17/01/P01037}{\emph{JINST} {\bfseries 17} (2022) P01037} [\href{https://arxiv.org/abs/2110.13961}{{\ttfamily 2110.13961}}].

\bibitem{Adams:2019uqx}
C.~Adams, M.~Del~Tutto, J.~Asaadi, M.~Bernstein, E.~Church, R.~Guenette et~al., \emph{{Enhancing neutrino event reconstruction with pixel-based 3D readout for liquid argon time projection chambers}}, \href{https://doi.org/10.1088/1748-0221/15/04/P04009}{\emph{JINST} {\bfseries 15} (2020) P04009} [\href{https://arxiv.org/abs/1912.10133}{{\ttfamily 1912.10133}}].

\bibitem{Spieler2005-pv}
H.~Spieler, \emph{Semiconductor Detector Systems}, Series on Semiconductor Science and Technology, Oxford University Press, London, England (Oct., 2005).

\bibitem{Dwyer:2018phu}
D.A.~Dwyer et~al., \emph{{LArPix: Demonstration of low-power 3D pixelated charge readout for liquid argon time projection chambers}}, \href{https://doi.org/10.1088/1748-0221/13/10/P10007}{\emph{JINST} {\bfseries 13} (2018) P10007} [\href{https://arxiv.org/abs/1808.02969}{{\ttfamily 1808.02969}}].

\bibitem{DUNE:2021tad}
{\scshape DUNE} collaboration, \emph{{Deep Underground Neutrino Experiment (DUNE) Near Detector Conceptual Design Report}}, \href{https://doi.org/10.3390/instruments5040031}{\emph{Instruments} {\bfseries 5} (2021) 31} [\href{https://arxiv.org/abs/2103.13910}{{\ttfamily 2103.13910}}].

\bibitem{Q-Pix:2022zjm}
{\scshape Q-Pix} collaboration, \emph{{Enhanced low-energy supernova burst detection in large liquid argon time projection chambers enabled by Q-Pix}}, \href{https://doi.org/10.1103/PhysRevD.106.032011}{\emph{Phys. Rev. D} {\bfseries 106} (2022) 032011} [\href{https://arxiv.org/abs/2203.12109}{{\ttfamily 2203.12109}}].

\bibitem{Miao:2023ivo}
P.~Miao, J.~Asaadi, J.B.R.~Battat, M.~Han, K.~Keefe, S.~Kohani et~al., \emph{{Demonstrating the Q-Pix front-end using discrete OpAmp and CMOS transistors}},  \href{https://arxiv.org/abs/2311.09568}{{\ttfamily 2311.09568}}.

\bibitem{Sauli2022}
F.~Sauli, \emph{Gaseous Radiation Detectors: Fundamentals and Applications}, Cambridge University Press (Nov., 2022), \href{https://doi.org/10.1017/9781009291200}{10.1017/9781009291200}.

\bibitem{Rolandi2008}
L.~Rolandi, W.~Riegler and W.~Blum, \emph{Particle Detection with Drift Chambers}, Springer Berlin Heidelberg (2008), \href{https://doi.org/10.1007/978-3-540-76684-1}{10.1007/978-3-540-76684-1}.

\bibitem{AlAtoum2020}
B.A.~Atoum, S.~Biagi, D.~Gonz{\'{a}}lez-D{\'{\i}}az, B.~Jones and A.~McDonald, \emph{Electron transport in gaseous detectors with a python-based monte carlo simulation code}, \href{https://doi.org/10.1016/j.cpc.2020.107357}{\emph{Computer Physics Communications} {\bfseries 254} (2020) 107357}.

\bibitem{cernGEM}
``Gaseous electron multipler at {CERN}.'' \url{https://gdd.web.cern.ch/gem}.

\bibitem{kevinThesis}
K.~Keefe, \emph{Development of Digital Architectures for Pixelated Readout of Time Projection Chambers: Q-Pix}, Ph.D. thesis, University of Hawaii, 2023.

\bibitem{leicester1984}
R.~Skelding, \emph{Electron transport coefficients in gas mixtures}, Ph.D. thesis, University of Leicester, 1984.

\bibitem{Battat2014}
J.B.~Battat, C.~Deaconu, G.~Druitt, R.~Eggleston, P.~Fisher, P.~Giampa et~al., \emph{The dark matter time projection chamber 4shooter directional dark matter detector: Calibration in a surface laboratory}, \href{https://doi.org/10.1016/j.nima.2014.04.010}{\emph{Nuclear Instruments and Methods in Physics Research Section A: Accelerators, Spectrometers, Detectors and Associated Equipment} {\bfseries 755} (2014) 6–19}.

\end{thebibliography}\endgroup

\end{document}